\documentclass[prl,twocolumn,floatfix]{revtex4}

\usepackage{amssymb}
\usepackage{graphicx}
\usepackage{amsmath}
\usepackage{amsfonts}

\begin{document}

\title{Towards Fault Tolerant Adiabatic Quantum Computation}
\author{Daniel A. Lidar}
\affiliation{Departments of Chemistry, Electrical Engineering, and Physics, Center for
Quantum Information Science \& Technology, University of Southern
California, Los Angeles, CA 90089}

\begin{abstract}
I show how to protect adiabatic quantum computation (AQC) against
decoherence and certain control errors, using a hybrid methodology involving
dynamical decoupling, subsystem and stabilizer codes, and energy gaps. 
Corresponding error bounds are derived. As an example I show how to perform
decoherence-protected AQC against local noise using at most two-body
interactions.
\end{abstract}

\pacs{03.67.Lx,03.67.Pp}
\maketitle

Adiabatic quantum computation (AQC), originally developed to solve
optimization problems \cite{Farhi:00}, offers a fascinating alternative to
the standard circuit model \cite{Nielsen:book} to which it is
computationally equivalent \cite{Aharonov:04Kempe:04Siu:04MLM:06}. The
effects of decoherence on AQC\ were studied in several works \cite%
{Childs:01,SarandyLidar:05,RolandCerf:04Tiersch:07}. Unlike the circuit
model, for which an elaborate theory of fault tolerant QC exists along with
a noise threshold for fault tolerance \cite{Aliferis:07}, it is not yet
known how to make AQC fault tolerant. Here I show how AQC can be can be
protected against decoherence and certain control errors. To do so I devise
a hybrid method that involving dynamical decoupling (DD) \cite%
{Zanardi:98bViola:99}, subsystem \cite{Viola:00a,Zanardi:99d,Kempe:00} and
stabilizer codes \cite{Gottesman:97a}, and energy gaps \cite%
{Bacon:01Weinstein:07,Jordan:05}.

Viewed as a closed system, AQC proceeds via slow evolution on a timescale
set by the system's minimal energy gap $\Delta $ from the ground state \cite%
{Farhi:00,Aharonov:04Kempe:04Siu:04MLM:06}. In the presence of the
system-bath interaction $H_{SB}$ this gap can be significantly reduced
because the interaction will cause energy level splittings, or an effective
broadening of system energy levels; when these levels overlap adiabaticity
breaks down and so does AQC, even at zero temperature \cite{SarandyLidar:05}%
. A bath at finite temperature presents another problem: in the universality
proofs \cite{Aharonov:04Kempe:04Siu:04MLM:06} the system energy gap scales
as an inverse polynomial in the problem size, so that the temperature too
must be lowered polynomially to prevent thermal excitations. All of the
problems listed above are due to the presence of $H_{SB}$. Clearly, if $%
H_{SB}$ can be effectively eliminated or reduced, this will enhance the
fidelity of AQC. The main tool I shall use to this end is dynamical decoupling, which
involves the application of strong and fast pulses. Perhaps
surprisingly, this can be done without interfering with the slow
adiabatic evolution.

\textit{Distance measure and operator norm}.--- As a distance measure
between states I use the trace distance $D[\rho _{1},\rho_{2}]\equiv \frac{1%
}{2}\|\rho _{1}-\rho _{2}\|_{1}$, where $\|A\|_{1}\equiv \mathrm{Tr}|A|$, $%
|A|\equiv \sqrt{A^{\dag }A}$ \cite{Nielsen:book}. When applied to pure
states $\rho _{i}=|\psi _{i}\rangle \langle \psi _{i}|$ I shall write $%
D[\psi _{1},\psi _{2}]$. The operator norm is $\|A\|\equiv \sup_{\left\Vert
|\psi \rangle \right\Vert =1}\left\Vert A|\psi \rangle \right\Vert
=\max_{i}\lambda _{i}$, where $\lambda _{i}\in \mathrm{Spec}(|A|)$.

\textit{Closed-system adiabatic error}.--- Let $s=t/T\in \lbrack 0,1]$ be
the dimensionless time, with $T$ the final time. Let the system Hamiltonian
that implements AQC, $H_{\mathrm{ad}}(s)$, act on $n$ qubits. In AQC the
ground state $|\phi _{\mathrm{ad}}(s)\rangle $ of $H_{\mathrm{ad}}(s)$ at
the final time $s=1$ encodes the solution to the computational problem \cite%
{Farhi:00}. The actual final state $|\psi (1)\rangle $ is the solution of
the Schr\"{o}dinger equation $d|\psi \rangle /ds=-iTH_{\mathrm{ad}}|\psi
\rangle $ ($\hbar =1$ units are used throughout). In AQC one is therefore
interested in minimizing the error $\delta _{\mathrm{ad}}\equiv D[\psi
(1),\phi _{\mathrm{ad}}(1)]$. Most of the known AQC algorithms interpolate
between initial and final local Hamiltonians, $H_{0}$ and $H_{1}$, via $H_{%
\text{\textrm{ad}}}(s)=(1-f(s))H_{0}+f(s)H_{1}$, where $f(0)=0$ and $f(1)=1$%
, and exhibit a final time that scales as a polynomial in the problem/system
size $n$. Locality means that $\|H_{\text{\textrm{ad}}}\|\sim \Delta _{0}O(n)
$, where $\Delta _{0}$ is the energy scale. Thus $\|d^{j}H_{\text{\textrm{ad}%
}}/ds^{j}\|\sim \Delta _{0}|d^{j}f/ds^{j}|O(n)$. Let $\{E_{i}(s)\}_{i=0}$ be
the eigenvalues of $H_{\mathrm{ad}}(s)$, and let $\Delta \equiv
\min_{i,s}|E_{i}(s)-E_{0}(s)|$ be the minimum gap from the
instantaneous ground state energy $E_0(s)$. Assume that $\Delta
(n)\sim \Delta 
_{0}n^{-z}$, where $z>0$ is the dynamical critical exponent. Depending on
the differentiability of $H_{\mathrm{ad}}$, and assuming that $\dot{H}_{%
\mathrm{ad}}(0)=\dot{H}_{\mathrm{ad}}(1)=0$, one can prove different
versions of the adiabatic theorem. For example, (i) \cite{Jansen:06}:\ if $%
H_{\mathrm{ad}}(s)$ is twice differentiable on $[0,1]$ then provided $T\sim
r\Vert \dot{H}_{\mathrm{ad}}\Vert ^{2}/\Delta ^{3}$ the error can be made
arbitrarily small in the time dilation factor $r>1$: $\delta _{\mathrm{ad}%
}<r^{-2}$. Or, (ii) \cite{HL:08}: if $H_{\mathrm{ad}}(s)$ is infinitely
differentiable on $[0,1]$ then provided $T \sim rN\Vert \dot{H}_{\mathrm{ad}%
}\Vert /\Delta ^{2}$, the error can be made exponentially small in the order 
$N$ of an asymptotic expansion: $\delta _{\mathrm{ad}}<r^{-N}$. In both
cases 
\begin{equation}
T\sim n^{\zeta }/\Delta _{0},  \label{eq:T}
\end{equation}%
where $\zeta =3z+2$ for case (i)\ and $\zeta =2z+1$ for case (ii), and I
omitted $|d^{j}f/ds^{j}|$. In AQC the interpolation from $H_{\mathrm{ad}}(0)$
to $H_{\mathrm{ad}}(1)$ can be chosen at will, in particular so as to
satisfy the above conditions on $H_{\mathrm{ad}}$. This shows that
closed-system AQC is resilient against control errors which cause $H_{%
\mathrm{ad}}(s)$ to deviate from its intended path, as long as these do not
modify the end point $H_{\mathrm{ad}}(1)$. This is a form of inherent fault
tolerance to control errors which is not shared by the circuit model
\cite{comment-AQCDD}. 

\textit{Open system evolution}.--- A description in terms of $H_{\mathrm{ad}%
} $ alone neglects the fact that in reality the adiabatic quantum computer
system is never perfectly isolated. The actual Hamiltonian is $%
H(t)=H_{S}(t)\otimes {I}_{B}+{I}_{S}\otimes H_{B}+H_{SB}$, where ${I}$
denotes the identity operator, $H_{S}=H_{\mathrm{ad}}+H_{\mathrm{C}}$ ($%
H_{B} $) acts on the system (bath) alone, $H_{\mathrm{C}}(t)$ is a control
Hamiltonian, and $H_{SB}=\sum_{\alpha }S_{\alpha }\otimes B_{\alpha }$,
where $S_{\alpha }$ ($B_{\alpha }$) acts on the system (bath). The role of $%
H_{\mathrm{C}}$ is to implement a DD procedure. The total propagator is $%
U(t)=\mathcal{T}\exp [-i\int_{0}^{t}H(t^{\prime })dt^{\prime }\,]$, where $%
\mathcal{T}$ denotes time ordering. The time evolved system state is $\rho
_{S}(t)=\mathrm{Tr}_{B}\rho (t)$, where $\rho (t)=U(t)\rho (0)U(t)^{\dag }$
is the joint system-bath state. Below I explain how to choose $H_{\mathrm{C}%
}(t)$ so that%
\begin{equation}
\lbrack H_{\mathrm{ad}}(t),H_{\mathrm{C}}(t^{\prime })]=0\quad \forall
t,t^{\prime }.  \label{eq:commute}
\end{equation}%
It is this condition that will allow application of DD without interfering with
the adiabatic evolution. Consider the \emph{uncoupled} setting
$H_{SB}=0$, to be
denoted by the superscript $0$. The ideal, noise-free
adiabatic system state is $\rho _{S,\mathrm{ad}}^{0}(t)=|\phi _{\mathrm{ad}%
}(t)\rangle \langle \phi _{\mathrm{ad}}(t)|$. Because the adiabatic,
control, and bath Hamiltonians all commute we have $\rho ^{0}(t)=\rho
_{S}^{0}(t)\otimes \rho _{\mathrm{C}}^{0}(t)\otimes \rho _{B}^{0}(t)$, where 
$\rho _{S}^{0}(t)=|\psi (t)\rangle \langle \psi (t)|$ [$\rho _{\mathrm{C}%
}^{0}(t)=|\psi _{\mathrm{C}}(t)\rangle \langle \psi _{\mathrm{C}}(t)|$] is
the actual system evolution under $H_{\mathrm{ad}}$ [$H_{\mathrm{C}}$], and $%
\rho _{B}^{0}(t)$ is the bath state evolved under $H_{B}$. Let
$\rho _{\mathrm{ad}}^{0}(t)\equiv \rho
_{S,\mathrm{ad}}^{0}(t)\otimes \rho _{\mathrm{C}}^{0}(t)\otimes \rho
_{B}^{0}(t)$ denote the ``ideal adiabatic joint state,'' with purely
adiabatic evolution of the first factor. Note that  
$\rho _{S}^{0}(0)=\rho _{S,\mathrm{ad}}^{0}(0)$.

\textit{General error bound}.--- Let $d$ ($\delta $) denote distances
in the joint (system) Hilbert space. To quantify the deviation
of the actual evolution from the desired one, let: 
\begin{equation*}
\begin{tabular}{ll}
$\delta _{S}\equiv D[\rho _{S}(T),\rho _{S,\mathrm{ad}}^{0}(T)],$ & $d_{%
\mathrm{D}}\equiv D[\rho (T),\rho ^{0}(T)]$ \\ 
$d_{\mathrm{ad}}\equiv D[\rho ^{0}(T),\rho _{\mathrm{ad}}^{0}(T)]=\delta _{%
\mathrm{ad}},$ & $d_{\mathrm{tot}}\equiv D[\rho (T),\rho _{\mathrm{ad}%
}^{0}(T)].$%
\end{tabular}%
\end{equation*}%
The overall objective is to minimize the distance $\delta
_{S}$ between the actual system state and the ideal, noise-free adiabatic
system state. The distance between the uncoupled joint state and the ideal
adiabatic joint state is $d_{\mathrm{ad}}$, which equals $\delta _{\mathrm{ad%
}}$ since $\|A\otimes B\|_{1}=\|A\|_{1}\|B\|_{1}$ and $\|\rho
_{B}^{0}\|_{1}=\|\rho _{\mathrm{C}}^{0}\|_{1}=1.$ The \textquotedblleft
decoupling distance\textquotedblright\ is $d_{\mathrm{D}}$:\ the distance
between the joint state in the coupled and uncoupled settings. Minimization
of this distance is the target of the DD\ procedure. Finally, $d_{\mathrm{tot%
}}$ is the distance between the actual and ideal joint states.

Because taking the partial trace can only decrease the distance between
states \cite{Nielsen:book}, we have $\delta _{S}\leq d_{\mathrm{tot}}$.
Using the triangle inequality 
we have $d_{\mathrm{tot}}\leq d_{\mathrm{D}}+d_{\mathrm{ad}}$. Therefore:%
\begin{equation}
\delta _{S}\leq d_{\mathrm{D}}+\delta_{\mathrm{ad}}.  \label{eq:deltaS}
\end{equation}%
This key inequality shows that the total system error is bounded above by
the sum of two errors:\ (i) due to the system-bath interaction in the
presence of decoupling ($d_{\mathrm{D}}$); (ii) due to the deviations from
adiabaticity in the \emph{closed} system ($d_{\mathrm{ad}}$). I shall
present a procedure intended to minimize $d_{\mathrm{D}}$ jointly with
$d_{\mathrm{ad}}$. 
This is an optimization problem: generically decoherence (closed-system
adiabaticity) worsens (improves) with increasing $T$.

\textit{Dynamical decoupling}.--- I now show how to minimize the decoupling
error $d_{\mathrm{D}}$. To do so I propose to apply strong and fast
dynamical decoupling (DD)\ pulses to the system on top of the adiabatic
evolution. It is convenient to first transform to an interaction picture
defined by $H_{\mathrm{ad}}+H_{B}$, i.e., $U(t)=U_{\mathrm{ad}}(t)\otimes
U_{B}(t)\tilde{U}(t)$, where $U_{X}(t)=\mathcal{T}\exp
[-i\int_{0}^{t}H_{X}(t^{\prime })dt^{\prime }\,]$, $X\in \{\mathrm{ad},B\}$.
Then $\tilde{U}$ satisfies the Schr\"{o}dinger equation $\partial \tilde{U}%
/\partial t=-i\tilde{H}\tilde{U}$, with $\tilde{H}=U_{B}^{\dag }\otimes U_{%
\mathrm{ad}}^{\dag }[H_{\mathrm{C}}+H_{SB}]U_{B}\otimes U_{\mathrm{ad}}=H_{%
\mathrm{C}}+\tilde{H}_{SB}$, where the second equality required Eq.~(\ref%
{eq:commute}). Define an effective \textquotedblleft error
Hamiltonian\textquotedblright\ $H_{\mathrm{eff}}(t)$ via $\tilde{U}%
(t)=e^{-itH_{\mathrm{eff}}(t)}$, which can be conveniently evaluated using
the Magnus expansion \cite{Casas:07}. Now consider a sequence of
non-overlapping control Hamiltonians $H_{\mathrm{DD}}^{(k)}(t)$ applied for
duration $w$ (pulse width) at pulse intervals $\tau $, i.e., $H_{\mathrm{C}%
}(t)=0$ for $t_{k}\leq t<t_{k+1}-w$ and $H_{\mathrm{C}}(t)=H_{\mathrm{DD}%
}^{(k)}$ for $t_{k+1}-w\leq t<t_{k+1}$, where $t_{k}=k(\tau +w)$, $k\in 
\mathbb{Z}_{K}$. The sequence $\{H_{\mathrm{DD}}^{(k)}\}_{k=0}^{K-1}$
defines a \textquotedblleft DD\ protocol\textquotedblright\ with cycle time $%
T_{c}=K(\tau +w)$ and unitary pulses $P_{k}$ generated by $\tilde{H}(t)=H_{%
\mathrm{DD}}^{(k)}+\tilde{H}_{SB}$, $t_{k+1}-w\leq t<t_{k+1}$. 
In the \textquotedblleft ideal pulse limit\textquotedblright\ $w=0$ one
defines the \textquotedblleft decoupling group\textquotedblright\ $\mathcal{G%
}=\{G_{k}\equiv P_{K-1}\cdots P_{k+1}P_{k}\}_{k=0}^{K-1}$ such that $%
G_{0}=I_{S}$. Then the total propagator becomes $\tilde{U}%
(T_{c})=\prod_{k=0}^{K-1}\exp [-i\tau (G_{k}^{\dag }\tilde{H}%
_{SB}G_{k})]\equiv e^{-iT_{c}H_{\mathrm{eff}}^{\mathrm{id}}}$, where $H_{%
\mathrm{eff}}^{\mathrm{id}}$ denotes the resulting effective Hamiltonian,
with Magnus series $H_{\mathrm{eff}}^{\mathrm{id}}=\sum_{j=0}^{\infty }H_{%
\mathrm{eff}}^{\mathrm{id}(j)}$ \cite{Zanardi:98bViola:99}. To lowest order:

\begin{equation}
H_{\mathrm{eff}}^{\mathrm{id}(0)}= \frac{1}{K}\sum_{k=0}^{K-1}G_{k}^{\dag }%
\tilde{H}_{SB}G_{k} \equiv \Pi_\mathcal{G}(\tilde{H}_{SB}).
\end{equation}%
In the limit $\tau \rightarrow 0$ one has $H_{\mathrm{eff}}^{\mathrm{id}}=H_{%
\mathrm{eff}}^{\mathrm{id}(0)}$, so that by properly choosing $\mathcal{G}$
one can effectively eliminate $H_{SB}$.

Returning to non-ideal ($w>0$) pulses, we have shown
by use of $\|[A,B]\|_{1}\leq 2\|A\|\|B\|_{1}$ and the Dyson expansion
that minimization of the \textquotedblleft error
phase\textquotedblright\ $\Phi (T)\equiv T\|H_{\mathrm{eff}}(T)\|$ implies
minimization of the decoupling distance $d_{\mathrm{D}}$
\cite{LZK:08}:
\begin{eqnarray}
  d_{\mathrm{D}} &\leq& \min [1,(e^{\Phi }-1)/2] \notag \\
  &\leq& \Phi \text{\quad if }\Phi
\leq 1.  \label{eq:D-Phi}
\end{eqnarray}%
For single-qubit systems we and others have shown that concatenated DD pulse
sequences can decrease $\Phi $ exponentially in the number of concatenation
levels \cite{KL1KL2Yao:07Witzel:07Zhang:08}. Here I focus on periodic pulse
sequences for simplicity. In periodic DD (PDD) one repeatedly applies the
DD\ protocol $\{H_{\mathrm{DD}}^{(k)}\}_{k=0}^{K-1}$ to the system, i.e., $%
H_{\mathrm{C}}(t+lK)=H_{\mathrm{C}}(t)$, $l\in \mathbb{Z}_{L}$. The total
time is thus $T=L(\tau +w)$, where the total number of pulses is $L$ and the
number of cycles is $L/K$. A calculation of the total error phase $\Phi (T)$
proceeds in two steps. First we find an upper bound $\Theta _{l}$ on $\Phi
_{l}(T_{c})$ for the $l$th cycle, using the Magnus expansion. Then we upper
bound $\Phi (T)$ by $(L/K)\max_{l}\Theta _{l}$. Let $J\equiv \|H_{SB}\|$
(system-bath coupling strength), $\beta \equiv \|H_{\text{\textrm{ad}}%
}+H_{B}\|\leq \beta _{S}+\beta _{B}$, where $\beta _{S}=\|H_{\text{\textrm{ad%
}}}\|$ and $\beta _{B}=\|H_{B}\|$, and $\alpha =O(1)$ a constant. 
A worst case analysis yields 
\cite{KhodjastehLidar:08}: 
\begin{equation}
\Phi (T)\leq \frac{\alpha (JT)^{2}}{L/K}+\frac{JTw}{\tau +w}+JT(\frac{\exp
(2\beta T_{c})-1}{2\beta T_{c}}-1),  \label{eq:Phi-bound}
\end{equation}%
This bound is valid as long the third term is $\leq JT$ and the Magnus
series is absolutely convergent over each cycle, a sufficient condition for
which is $JT_{c}<\pi $ \cite{KhodjastehLidar:08,Casas:07}.

\textit{Joint AQC-DD optimization}.--- Recall Eq.~(\ref{eq:T}) for closed
system adiabaticity. The given and fixed parameters of the problem are $J$, $%
\Delta _{0}$, and $z$ (or $\zeta $). The task is to ensure that each of the
terms in Eq. (\ref{eq:Phi-bound}) vanishes as a function of $n$. I
show in \cite{comment2} that if $\tau $ and $w$ scale as 
\begin{equation}
\tau \sim n^{-(\zeta +\epsilon _{1})}/\Delta _{0},\qquad w\sim n^{-(2\zeta
  +\epsilon _{1}+\epsilon _{2})}/J,
\label{eq:tau-w}
\end{equation}%
with $\epsilon _{1}>1$ and $\epsilon _{2}>0$, then 
\begin{equation}
d_{\mathrm{D}}\lesssim (J/\Delta _{0})^{2}n^{-\epsilon _{1}}+n^{-\epsilon
_{2}}+(J/\Delta _{0})n^{1-\epsilon _{1}},  \label{eq:D_D}
\end{equation}%
which is arbitrarily small in the large $n$ limit. Combining this with the
bounds above ($\delta _{\mathrm{ad}}<r^{-2}$
or $\delta _{\mathrm{ad}}<r^{-N}$) and inequality
(\ref{eq:deltaS}), it follows that
for an AQC algorithm with time scaling as $T=L(\tau +w)\sim \Delta
_{0}^{-1}n^{\zeta }$, the total error $\delta _{S}$ can be made arbitrarily
small. This is the first main result of this work: \emph{using PDD with properly chosen
parameters we can obtain arbitrarily accurate AQC}.

However, there is a shortcoming: the pulse
intervals and widths must shrink with $n$ as a power law, with an exponent
dictated by the dynamical critical exponent $z$ of the model [Eq.~(\ref{eq:tau-w})]. I expect that
this can be remedied by employing concatenated DD
\cite{KL1KL2Yao:07Witzel:07Zhang:08,KhodjastehLidar:08}.

\textit{Seamless AQC-DD}.--- The entire analysis relies so far on the
\textquotedblleft non-interference\textquotedblright\ condition (\ref%
{eq:commute}). When can it be satisfied? Fortunately, the general background
theory was worked out in \cite{Viola:00a,Zanardi:99d}, though without any
reference to AQC. I review this theory and make the connection to AQC
explicit. The decoupling group $\mathcal{G}$ induces a decomposition of the
system Hilbert space $\mathcal{H}_{S}$ via its group algebra $\mathbb{C}%
\mathcal{G}$ and its commutant $\mathbb{C}\mathcal{G}^{\prime }$, as follows:%
\begin{align}
\mathcal{H}_{S}& \cong \bigoplus_{J}\mathbb{C}^{n_{J}}\otimes \mathbb{C}%
^{d_{J}},  \label{eq:Hsplit} \\
\mathbb{C}\mathcal{G}& \cong \bigoplus_{J}I_{n_{J}}\otimes M_{d_{J}},\quad 
\mathbb{C}\mathcal{G}^{\prime }\cong \bigoplus_{J}M_{n_{J}}\otimes I_{d_{J}}.
\end{align}%
Here $n_{J}$ and $d_{J}$ are, respectively, the multiplicity and dimension
of the $J$th irreducible representation (irrep) of the unitary
representation chosen for $\mathcal{G}$, while $I_{N}$ and $M_{N}$ are,
respectively, the $N\times N$ identity matrix and unspecified complex-valued 
$N\times N$ matrices. The adiabatic state is encoded into (one of) the left
factors $\emph{C}_{J}\equiv \mathbb{C}^{n_{J}}$, i.e., each such factor
(with $J$ fixed) represents an $n_{J}$-dimensional code $\emph{C}_{J}$
storing $\log _{d}n_{J}$ qu$d$its. The DD pulses act on the right factors. 
As shown in \cite{Viola:00a}, the dynamically decoupled evolution on each
factor (code) $\emph{C}_{J}$ will be noiseless in the ideal limit $w,\tau
\rightarrow 0$ iff $\Pi _{\mathcal{G}}(S_{\alpha })=\bigoplus_{J}\lambda
_{J,\alpha }I_{n_{J}}\otimes I_{d_{J}}$ for all system operators $S_{\alpha
} $ in $H_{SB}$, whence $H_{\mathrm{eff}}^{\mathrm{id}(0)}=\bigoplus_{J}%
\left[ \left( I_{n_{J}}\otimes I_{d_{J}}\right) \right] _{S}\otimes \left[
\sum_{\alpha }\lambda _{J,\alpha }B_{\alpha }\right] _{B}$. Thus, assuming
the latter condition is met, \emph{under the action of DD the action of }$H_{%
\mathrm{eff}}^{\mathrm{id}(0)}$ \emph{on the code} $\emph{C}_{J}$\emph{\ is
proportional to }$I_{n_{J}}$\emph{, i.e., is harmless}. Quantum logic, or
AQC, is enacted by the elements of $\mathbb{C}\mathcal{G}%
^{\prime }$. Dynamical decoupling operations are enacted via the elements of 
$\mathbb{C}\mathcal{G}$. \emph{Condition} (\ref{eq:commute}) 
\emph{is satisfied because} $[\mathbb{C}\mathcal{G},\mathbb{C}\mathcal{G}^{\prime }]=0$.

\textit{Stabilizer decoupling}.--- An important example of the general $%
\mathbb{C}\mathcal{G}/\mathbb{C}\mathcal{G}^{\prime }$construction is when $%
\mathcal{G}$ is the stabilizer of a quantum error correcting code and the
commutant is the normalizer $\mathcal{N}$ of the code \cite{Gottesman:97a}.
Because a stabilizer group is Abelian its irreps are all one-dimensional. A
stabilizer code encoding $n$ qubits into $n_{J}=k$ has $n-k$ generators,
each of which has eigenvalues $\pm 1$. Then $J$ runs over the $2^{n-k}$
different binary vectors of eigenvalues, meaning that $\mathcal{H}_{S}\cong
\bigoplus_{J=\{\pm 1,...,\pm 1\}}\mathbb{C}^{2^{k}}$, and each of the
subspaces in the sum is a valid code $\emph{C}_{J}$. Here the
elements of $\mathcal{N}$ are viewed as Hamiltonians. For this reason 
only the encoded single-qubit normalizer operations are required; encoded
two-body interactions are constructed as tensor products of single-qubit
ones.

\textit{Energy-gap protection}.--- Application of DD pulses is the main
mechanism I propose for protection of AQC, but it has a shortcoming as
noted above. Fortunately, the formulation presented here easily accommodates the AQC
energy-gap protection strategy proposed in \cite{Jordan:05}, which can
be viewed as adding another layer of protection for dealing with
finite-resource-DD. Namely, if the decoupling group $\mathcal{G}$ is also a
stabilizer group for code $\emph{C}_{J}$, then for each Pauli error $%
S_{\alpha }$ in $H_{SB}$ there is at least one element $P_{j}\in \mathcal{G}$
such that $\{P_{j},S_{\alpha }\}=0$, and otherwise $[P_{j},S_{\alpha }]=0$ 
\cite{Gottesman:97a}. We can then add an energy penalty term $H_{\mathrm{P}%
}=-E_{\mathrm{P}}\sum_{j=1}^{|\mathcal{G}|-1}P_{j}\in \mathbb{C}\mathcal{G}$
to $H_{S}$, where $E_{\mathrm{P}}>0$ is the penalty. Imperfect decoupling
means that $H_{\mathrm{eff}}^{\mathrm{id}(j\geq 1)}\neq 0$. 
To lowest order, $H_{\mathrm{eff}}^{\mathrm{id}(1)}=\sum_{\alpha }S_{\alpha
}\otimes B_{\alpha }^{(1)}$, and an \textquotedblleft erred
state\textquotedblright\ will be of the form $|\psi _{\alpha }^{\bot
}\rangle =S_{\alpha }|\psi \rangle $, where $|\psi \rangle =P_{j}|\psi
\rangle \in \emph{C}_{J}$ $\forall j$. Then $H_{\mathrm{P}}|\psi _{\alpha
}^{\bot }\rangle =%
\left\{ \left[ a-(K-1)\right] (K-1)E_{\mathrm{P}}\right\} |\psi _{\alpha
}^{\bot }\rangle $, where $a$ is the number of stabilizer elements that
anticommute with $S_{\alpha }$. Thus $|\psi _{\alpha }^{\bot }\rangle $ is
an eigenstate of $H_{\mathrm{P}}$ and has $a(K-1)E_{\mathrm{P}}$ more energy
than any state in the code space. Ref.~\cite{Jordan:05} showed, using a
Markovian model of qubits coupled to a photon bath, the important result
that this energy gap for erred states implies that the temperature need only
shrink logarithmically rather than polynomially in the problem size.
However, note that to deal with generic system-bath interactions both the
stabilizer and normalizer elements must involve $k$-local interactions, with 
$k>2$ \cite{Jordan:05}.

\textit{2-local decoherence-resistant universal AQC}.--- First recall a
recent universality result. The following simple 2-local Hamiltonian allows
for universal AQC \cite{Biamonte:07}: $H_{\mathrm{ad}}^{\mathrm{univ}%
}(t)=\sum_{i;\alpha \in \{x,z\}}h_{i}^{\alpha }(t)\sigma _{i}^{\alpha
}+\sum_{i,j;\alpha \in \{x,z\}}J_{ij}^{\alpha }(t)\sigma _{i}^{\alpha
}\sigma _{j}^{\alpha }$. With this all the tools have been assembled to
demonstrate the second main result of this work: a stabilizer decoupling procedure
against 1-local noise that uses only 2-local interactions. By 1-local noise
I mean the main nemesis of quantum computing, namely the linear decoherence
model: 
$H_{SB}^{\mathrm{lin}}=\sum_{\alpha =x,y,z}\sum_{j=1}^{n}\sigma _{j}^{\alpha
}\otimes B_{j}^{\alpha }$, 
where $\{B_{j}^{\alpha }\}$ are arbitrary bath operators. To beat $H_{SB}^{%
\mathrm{lin}}$, use the Abelian \textquotedblleft universal decoupling
group\textquotedblright\ \cite{Zanardi:98bViola:99} $\mathcal{G}_{\mathrm{uni%
}}=\{I,X,Y,Z\}$, where $X(Y,Z)=\bigotimes_{j=1}^{n}\sigma _{j}^{x(y,z)}$. It
is simple to verify that $\Pi _{\mathcal{G}_{\mathrm{uni}}}(H_{SB}^{\mathrm{%
\ lin}})=0$. 
As noted in Ref.~\cite{Viola:00a}, $\mathcal{G}_{\mathrm{uni}}$ is the
stabilizer of an $[[n,n-2,2]]$ stabilizer code $\mathcal{C}$,
whose codewords are $\{|\psi _{x}\rangle =\left( |x\rangle +|\mathrm{not\,}%
x\rangle \right) /\sqrt{2}\}$, where $x$ is an even-weight binary string of
length $n$, with $n$ even. For example, for $n=4$ we find: $|00\rangle
_{L}=\left( |0000\rangle +|1111\rangle \right) /\sqrt{2}$, $|10\rangle
_{L}=\left( |0011\rangle +|1100\rangle \right) /\sqrt{2}$, $|01\rangle
_{L}=\left( |0101\rangle +|1010\rangle \right) /\sqrt{2}$, $|11\rangle
_{L}=\left( |1001\rangle +|0110\rangle \right) /\sqrt{2}$. Now universal AQC
over $\mathcal{C}$ can be implemented using 2-local
Hamiltonians. To compute over $\mathcal{C}$ we replace each
Pauli matrix in $H_{\mathrm{ad}}^{\mathrm{univ}}$ by its encoded partner.
Encoded single-qubit operations for $\mathcal{C}$ are the
2-local\ $\bar{X}_{j}=\sigma _{1}^{x}\sigma _{j+1}^{x}$ and $\bar{Z}%
_{j}=\sigma _{j+1}^{z}\sigma _{n}^{z}$, where $j=1,...,n-2$. The 2-local
interactions $\sigma _{i}^{x}\sigma _{j}^{x}$ and $\sigma _{i}^{z}\sigma
_{j}^{z}$ appearing in $H_{\mathrm{ad}}$ are replaced by the 2-local $\bar{X}%
_{i}\bar{X}_{j}=\sigma _{i+1}^{x}\sigma _{j+1}^{x}$ and $\bar{Z}_{i}\bar{Z}%
_{j}=\sigma _{i+1}^{z}\sigma _{j+1}^{z}$. \emph{Thus we see that universal
AQC can be combined with DD using only 2-local }$\sigma _{i}^{x}\sigma
_{j}^{x}$ \emph{and} $\sigma _{i}^{z}\sigma _{j}^{z}$ \emph{interactions over%
} $\mathcal{C}$.

Examples of promising QC implementations where $X$, $Z$ (as pulses for DD)
and $\sigma _{i}^{x}\sigma _{j}^{x},\sigma _{i}^{z}\sigma _{j}^{z}$ (as
Hamiltonians for AQC) are available and controllable, are systems including
capacitive coupling of flux qubits \cite{Averin:03} and spin models
implemented with polar molecules \cite{Micheli:06}. Also note that in
principle, as discussed above, we can create an additional energy gap \cite%
{Jordan:05} against single-qubit errors by adding a penalty term $%
H_{P}=-E_{P}(X+Y+Z)$ to the system Hamiltonian. However, $H_{P}$ is an $n$%
-local interaction.

\textit{Conclusions and outlook}.--- Using a combination of various tools in
the arsenal of decoherence control I have shown how to protect AQC
against decoherence. While I believe that the methods proposed here should
significantly contribute towards the viability and robustness of AQC, what
is still missing is a threshold theorem for fault tolerant AQC. This will
most likely require the incorporation of feedback, in order to correct DD
pulse imperfections and other control noise \cite{comment-AQCDD}. One
possibility for doing so might be to perform syndrome measurements on the
commutant factor [$\mathbb{C}^{d_{J}}$ in Eq.~(\ref{eq:Hsplit})] as in
recent circuit-model fault tolerance work using subsystems codes \cite%
{Aliferis:07}.

\vspace{-.009cm} \textit{Acknowledgements}.--- Important discussions with K.
Khodjasteh, A. Hamma, and P. Zanardi are gratefully acknowledged. Supported
under grant NSF CCF-0523675.


\begin{thebibliography}{99}
\bibitem{comment2} To derive Eq. (\ref{eq:D_D}) let $L=Kn^{2\zeta +\epsilon
_{1}}$, and absorb $\alpha $ into $T$. The first term in Eq. (\ref%
  {eq:Phi-bound}) thus yields
$n^{-\epsilon _{1}}<(\Delta _{0}/J)^{2}$, which is satisfied for large
enough $n$ if (i) $\epsilon _{1}>0$. Using $T=L(\tau +w)$, the second term
yields $w<1/(JL)$, which we can satisfy with (ii) $w=n^{-(2\zeta +\epsilon
_{1}+\epsilon _{2})}/(JK)$, $\epsilon _{2}>0$, and the first term yields 
(iii) $w<1/(J\sqrt{LK})-\tau $; but $w>0$ so that together these imply (iv) $%
\tau <1/(J\sqrt{LK})$. We can satisfy (iii) and (iv) by choosing (v) $\tau
=n^{-(\zeta +\epsilon _{1})}/(K\Delta _{0})$, since then
$T=L(\tau +w)\overset{n\gg 1}{\rightarrow }\Delta _{0}^{-1}n^{\zeta }$ as it
should. As for the third term in Eq. (\ref{eq:Phi-bound}), note that $\frac{%
e^{x}-1}{x}-1\leq (e-2)x$ if $x\leq 1$. Indeed, $x=2\beta T_{c}\sim 2\beta
(n^{-(\zeta +\epsilon _{1})}/\Delta _{0}+n^{-(2\zeta +\epsilon _{1}+\epsilon
_{2})}/J)\ll 1$ for $\beta _{S} and \beta _{B}\sim O(n)$ (local Hamiltonians;
e.g., a scenario where each system particle is in contact with a local
(spin) bath). Therefore the third term is bounded above by 
$(e-2)2JT\beta T_{c}\overset{n\gg 1}{\rightarrow }O(n)(J/\Delta
_{0})n^{-\epsilon _{1}}$, so that we require $\epsilon _{1}>1$.
\end{thebibliography}

\end{document}